\documentclass[preprint,prb,aps,showkeys]{revtex4}
\usepackage{epsfig}
\usepackage{amsmath}
\begin{document}
\title{What ice can teach us about water interactions: a critical comparison of
the performance of different water models}
{\small Faraday Discussiosn, volume 141, 251-276, (2009)}
\author{C. Vega*, J. L. F. Abascal, M. M. Conde and J. L. Aragones}
\affiliation{Departamento de Qu\'{\i}mica F\'{i}sica,
Facultad de Ciencias Qu\'{\i}micas,\\ Universidad Complutense de Madrid,
28040 Madrid, Spain ,\\ 
*) cvega@quim.ucm.es}

\date{\today}

\begin{abstract}
The performance of several popular water models (TIP3P, TIP4P, TIP5P and 
TIP4P/2005) is analysed.
For that purpose the predictions for ten different properties of water are 
investigated, namely: 
1. vapour-liquid equilibria (VLE) and critical temperature; 2. surface tension; 
3. densities of the different solid structures of water (ices); 
4. phase diagram; 5. melting point properties; 6. maximum in 
 density at room pressure and thermal coefficients $\alpha$ and $\kappa_T$;
7. structure of liquid water and ice; 8. equation of state at high pressures; 
9. diffusion coefficient; 10. dielectric constant. 
For each property, the performance of each model is analysed in detail with a 
critical discussion of the possible reason of the success or failure of the 
model. 
A final judgement on the quality of these models is provided. 
TIP4P/2005 provides the best description of almost all properties of the list, 
with the only exception of the dielectric constant. 
In the second position, TIP5P and TIP4P yield an overall similar performance, 
and the last place with the poorest description of the water properties is 
provided by TIP3P.
The ideas leading to the proposal and design of the TIP4P/2005 are also 
discussed in detail. 
TIP4P/2005 is probably close to the best description of water that can be 
achieved with a non polarizable model described by a single Lennard-Jones
(LJ) site and three charges. 
\end{abstract}


\maketitle

\vspace{0.5cm}

\section{Introduction}

Water is probably the most important molecule in our relation to nature.
It forms the matrix of life,\cite{ball01} it is the most common solvent for
chemical processes, it plays a major role in the determination of the climate
on earth, and also it appears on planets, moons and comets.\cite{poirier82}
Water is interesting not only from a practical point of view, but also from
a fundamental point of view. 
In the liquid phase water presents a number of anomalies when compared to other
liquids.\cite{eisenberg69,bookPhysIce,finneyreview,chaplin,JML_2001_90_0303} 
In the solid phase it exhibits one of the most complex phase diagrams,
having fifteen different solid structures.\cite{eisenberg69} 
Due to its importance and its complexity, understanding the properties 
of water from a molecular point of view is of considerable interest. 
The experimental
study of the phase diagram of water has spanned the entire
20th century, starting with the pioneering works of Tammann\cite{tammann} and 
Bridgman\cite{bridgman12} up to the recent discovery of ices XII, XIII
and XIV.\cite{N_1998_391_00268,iceXIII} 
The existence of several types of amorphous phases at low temperatures, 
\cite{rice1,JPCM_2003_15_R1669,N_1984_310_00393} the possible existence of a liquid-liquid  phase transition
in water\cite{N_1992_360_00324_nolotengo,N_1998_396_00329} , the properties of ice at a free
surface\cite{kroes92,suter_jcp06,dash_2,salmeron,surface_melting_quantum,jedlovszky06,pnas_buch} and the 
interaction with hydrophobic molecules\cite{chandler2005,monson_jpcb07} have also been the focus of much interest in the last two decades.

Computer simulations of water started their road
with the pioneering papers by Watts and Barker\cite{CPL_1969_3_0144_nolotengo}
and by Rahman and Stillinger\cite{JCP_1971_55_03336} about forty years ago. 
A key issue when performing simulations of water is the choice of the potential
model used to describe the interaction
between molecules.\cite{lynden,laaksonen,ferrarioKF,paschek,waterinproteins} 
A number of different potential models have been proposed.
An excellent survey of the predictions of the different models proposed up 
to 2002 was made by Guillot.\cite{guillot02}.
Probably the general feeling is that no water potential model 
is totally satisfactory and that there are no significant differences 
between water models.

In the recent years the increase in the computing power has allowed
the calculation of new properties which can be used as new target quantities
in the fit of the potential parameters.
More importantly, some of these properties have revealed as a stringent
test of the water models.
In particular, recently we have determined  
the phase diagram for different water models and we have found that the 
performance of the models is quite different.\cite{sanz1,abascal05a,abascal05b} 
As a consequence, new potentials have been proposed.
In this paper we want to perform a detailed analysis of the performance
of several popular water models including the recently proposed TIP4P/2005, and
including in the comparison some properties of the solid phases, thus extending
the scope of the comparison performed by Guillot\cite{guillot02}. The importance
of including solid phase properties in the test of water models was advocated 
by Morse and Rice \cite{morse82} and  Whalley\cite{whalley84} , among others,
and Monson suggested that the same should be true for other substances\cite{ACP_2000_115_0113_nolotengo,monson08}.

It must be recognised that water is flexible and polarizable.
That should be taken into account in the next generation of water 
models.\cite{cummings05,siepmann2000} 
However, it is of interest at this stage to 
analyse how far it is possible to go in the description of real water with 
simple (rigid, non-polarizable) models.  
For this reason we will restrict our study to rigid non-polarizable models.
Besides, in case a certain model performs better than others, it would be of 
interest to understand why is this so since this information can be useful in 
the development of future models. 
As commented above, the study of the phase diagram of water can be extremely 
useful to discriminate among different water models. 
Thus, in the comparison between water models, we incorporate not only 
properties of liquid water but also properties of the solid phases of water.
The scheme of this paper is as follows. 
In Section II, the potential models used in this paper are described. 
In Section III, we present the properties that will be used 
in the comparison between the different water models. 
Section IV gives some details of the calculations of this work. 
Section V presents the results of this paper.
Finally, in Section VI, the main conclusions will be discussed. 

\section{Water potential models}
In this work we shall focus on rigid, non-polarizable models of water. 
All the models we are considering locate the positive charge on the hydrogen 
atoms and a Lennard-Jones interaction site on the position of the oxygen. 
What are the differences between them? 
They do differ in three significant aspects:
\begin{itemize}
\item {The bond geometry. By bond geometry we mean the choice of the OH bond 
 length and H-O-H bond angle of the model.}
\item {The charge distribution. All the models place one positive charge at the
 hydrogens but differ in the location of the negative charge(s).}
\item {The target properties. By target properties we mean the properties of 
 real water that were used to fit the potential parameters (forcing the model 
 to reproduce the experimental properties).}
\end{itemize}

In this paper we compare the performance of the following potential models:
TIP3P\cite{jorgensen83}, TIP4P\cite{jorgensen83}, TIP5P\cite{mahoney00}  and 
TIP4P/2005\cite{abascal05b}. 
This  selection presents an advantage. 
All these models exhibit the same bond geometry (the $d_{OH}$ distance and 
H-O-H angle in all these models are just those of the molecule in the gas phase,
namely 0.9572 \AA\ and 104.52 degrees, respectively). 
Therefore, any difference in their performance is 
due to the difference in their charge geometry and/or target properties. 
Let us now present each of these models. 

\subsection{TIP3P}
TIP3P was proposed by Jorgensen {\em et al.}\cite{jorgensen83}. 
In the TIP3P model the negative charge 
is located on the oxygen atom and the positive charge on the hydrogen atoms. 
The parameters of the model (i.e, the values of the LJ potential $\sigma$ and 
$\epsilon$ and the value of the charge on the hydrogen atom) were obtained by 
reproducing the vaporization enthalpy and liquid density of water at ambient
conditions.
TIP3P is the model used commonly in certain force fields of biological 
molecules.  
A model similar in spirit to TIP3P is SPC\cite{berendsen82}. 
We shall not discuss in this paper SPC model so extensively, since we want to 
keep the discussion within TIP$n$P geometry (by TIP$n$P geometry we mean that $d_{OH}$ 
adopts the value 0.9572 \AA\  and the H-O-H angle is 104.52 degrees).  
In the SPC model the OH bond length is 1 \AA\ and the H-O-H is 109.47 degrees 
(the tetrahedral value).
The parameters of SPC were obtained in the same way as those of TIP3P.
Thus TIP3P and SPC are very much similar models. 
Nine distances must be computed to evaluate the energy between two TIP3P (or SPC) water models so that their computational cost is proportional to 9.

\subsection {TIP4P}
The key feature of this model is that the site carrying the negative charge 
(usually denoted as the M site) is not located at the oxygen atom but on the 
H-O-H bisector at a distance of 0.15 \AA. 
This geometry was already proposed in 1933 by Bernal and Fowler.\cite{bernal33}.
TIP4P was proposed by Jorgensen {\em et al.}\cite{jorgensen83} who determined 
the parameters of the potential to also reproduce the vaporization enthalpy and
liquid density of liquid water at room temperature.  
It is fair to say that TIP4P, although quite popular, is probably less
often used than TIP3P or SPC/E. The reader may be surprised to learn that
the reasons for that are the appearance of a massless site (the M center)
and the apparently higher computational cost.
To deal with a massless site within a MD package one must either
solve the orientational equations of motion for 
a rigid body (for instance, using quaternions) or using constraints after 
distributing the  force acting on the M center among the rest of 
the atoms of the system. These options should be incorporated in the MD package 
and this is not always the case. Also the computational cost of TIP4P is proportional
to 16 when no trick is made (due to the presence of four interaction sites) but it
can be reduced to 10 by realizing that one must only compute 
the O-O distance to compute the LJ 
contribution, plus 9 distances to compute the Coulombic interaction.
Thus, for some users TIP4P is not a good choice as a water model either because
is slower than TIP3P in their molecular dynamics package or because the package
may have problems in dealing with a massless site.
These are methodological rather than scientifical reasons but they appear 
quite often. 
Some modern codes as GROMACS\cite{lindahl01} (just to mention one example) solve these two 
technical aspects 
quite efficiently.

\subsection {TIP5P}
TIP5P is a relatively recent potential:
it was proposed in 2000 by Mahoney and Jorgensen.\cite{mahoney00} 
This model is the modern version of the models used in the seventies (ST2) 
in which the negative charge was located at the position of the lone pair 
electrons.\cite{ST2} 
Thus instead of having one negative charge at the center M, this model has two 
negative charges at the L centres. 
Concerning the target properties this model reproduces the vaporization 
enthalpy and density of water at ambient conditions.
This is in common with TIP3P and TIP4P. 
However, TIP5P incorporates a new target property: the density maximum 
of liquid water. 
The existence of a maximum in density at approximately 277 K is one of the 
fingerprints properties of liquid water.
Obtaining density maxima in constant temperature-constant pressure (NpT) 
simulations of water models is not difficult 
from a methodological point of view.\cite{baez94,vega05b,paschek}
However, very long runs are required (of the order of several millions time 
steps (MD) or cycles (MC)) to determine accurately the location of the density
maximum.
Thus, it is not surprising that the very first model able to reproduce the 
density maximum was proposed in 2000 when the computing power allowed an 
accurate calculation of the temperature and density at the maximum.
Since TIP5P consists of five interaction sites, it apparently requires the 
evaluation of 25 distances. 
However, using the trick described for TIP4P, it is possible to show that 
one just requires the evaluation of 17 distances. 

\subsection {SPC/E}
In this paper we shall focus mainly on TIP$n$P like models. 
However, it is worth to introduce SPC/E,\cite{berendsen87} considered by many 
as one of the best water models.
SPC/E has the same bond geometry as SPC.
Concerning the charge distribution, it locates the negative charge at the 
oxygen atom, as SPC and TIP3P. 
The target properties of SPC/E were the density and the vaporization enthalpy 
at room temperature. 
So far everything seems identical to SPC. 
The key issue is that SPC/E only reproduces the vaporization enthalpy of real 
water when a polarization energy correction is included. 
Berendsen {\em et al.}\cite{berendsen87} pointed out that, when using 
non-polarizable models, one should include a polarization correction  
before comparing the vaporization enthalpy of the model to the experimental 
value. 
This is because the dipole moment is enhanced in these type of models in order
to account approximately for the neglected many body polarization forces.
Thus, when the vaporization enthalpy is calculated one compares the energy
of the liquid with that of a gas with the enhanced dipole moment.
It is then necessary to correct for the effect of the difference between the 
dipole moment of the isolated molecule and that of the effective dipole moment 
used for the condensed phase.
The correction term is given by:
\begin{equation}
\label{polarization}
  \frac{E_{pol}}{N} = \frac{(\mu-\mu_{gas})^{2}}{2\alpha_{p}}
\end{equation}
where $\alpha_p$ is the polarizability of the water molecule, $\mu_{gas}$ is 
the dipole moment of the molecule in the gas phase and $\mu$ is the dipole 
moment of the model. 
SPC/E reproduces the vaporization enthalpy of water only if the correction 
given by Eq.\ref{polarization} is used.
The introduction of the polarization correction is the essential feature of
SPC/E. 

\subsection {TIP4P/2005}
The TIP4P/2005 was recently proposed by Abacal and Vega\cite{abascal05b} after 
evaluating the phase diagram for the TIP4P and SPC/E models of 
water.\cite{sanz1,sanz2}
It was found that TIP4P provided a much better description of the
phase diagram of water than SPC/E. 
Both models yielded rather low melting points.\cite{vegafilosofico}
For this reason, it was clear that the TIP4P could be slightly modified, while
still predicting a correct phase diagram but improving the description of the 
melting point. 
The TIP4P/2005 has the same bond geometry as the TIP4P family. 
Besides TIP4P/2005 has the same charge distribution as TIP4P (although the 
distance from the M site to the oxygen atom is slightly modified).
The main difference between TIP4P and TIP4P/2005 comes from the choice of the 
target properties used to fit the potential parameters. 
TIP4P/2005 not only uses a larger number of target properties than any of 
the water models previously proposed but also they represent a wider range of
thermodynamic states.
As it has become traditional, the first of the target properties is the density
of water at room temperature and pressure (but in TIP4P/2005 we have also tried
to account for the densities of the solid phases).
More significantly, TIP4P/2005 incorporated as target properties those used 
in the succesfull SPC/E and TIP5P models. 
As in SPC/E, the polarization correction is included in the calculation of
the vaporization enthalpy.
Secondly (as in TIP5P) the maximum in density of liquid water at room pressure
(TMD) was used as a target property. 
Finally, completely new target properties are the melting temperature of
the hexagonal ice (Ih) and a satisfactory description of the complex region
of the phase diagram involving different ice polymorphs.

In Table \ref{potentials} the parameters of the models SPC, SPC/E, TIP4P, TIP5P
and TIP4P/2005 are shown.
The dipole moment and the three values of the traceless quadrupolar tensor are 
given in Table \ref{tabl_moments}. 
There, we also present the values of the quadrupole moment $Q_{T}$ which 
is defined as $Q_{T}= (Q_{xx}-Q_{yy})/2$. 
We have chosen the H-O-H bisector as the $Z$ axis, the $X$ axis in the 
direction of the line joining the hydrogen atoms so that the $y$ axis is
perpendicular to the molecular plane.
It can be shown that, for models consisting of just three charges, the value 
of $Q_{T}$ is independent of the origin (when the z axis is located along the H-O-H
bisector).\cite{abascal07a,abascal07b,abascal07c}
An interest feature appears in Table II. 
Although for most of water models the dipole moment is close to 2.3 Debyes,
they differ significantly in their values of the quadrupole moment. 
Notice that neither the dipole nor the quadrupole moment were used as target 
properties by any of the water models discussed here. 
Thus the charge distribution determines the aspect of the quadrupolar tensor. 
One may suspect that, since quadrupolar forces induce a strong orientational 
dependence, the differences in the quadrupolar tensor between different water 
models will be manifested significantly in the solid phases where the relative
orientation of the molecules is more or less fixed by the structure of the 
solid. 
The coupling between dipolar and quadrupolar interactions is
well known since some time ago due to early studies about the behavior of hard
spheres with dipole and quadrupole moments.\cite{patey76}
However, the role of the quadrupole in the properties of water has been 
probably overlooked although there were some clear warnings about its 
importance.\cite{finney85,watanabe89,tribello06}  

\section {An exam for water models}
Guillot presented in 2002\cite{guillot02} a study of the performance of
water models to describe several properties of water.
Although this is only six years ago 
there are several reasons to perform this study once again. 
At that time TIP5P was just released. TIP4P/2005 was proposed three years
later. 
In these years a more precise determination of some properties of the  water 
models (surface tension, temperature of maximum density) has been obtained.
More importantly, the calculation in the last years of water properties that 
were almost completely unknown before (as, for instance, predictions for the 
solid phases of water, the determination of the melting points and the phase 
diagram calculations) make interesting a new comparative of the performance
of the models.
It is necessary to select some properties to establish the comparison. 
The selected properties should be as many as possible but representative of 
the different research communities with interest in water. 
We will not include in the comparison properties of the gas phase (second 
virial coefficients, vapour densities) because only a polarizable model can be
successful in describing all the phases of water. 
Non-polarizable models can not describe simultaneously the vapour phase and the
condensed phases.
Thus, the models described above fail in describing vapour properties 
because they ignore the existence of the molecular polarizability.%
\cite{kofke_b2_water_1,kofke_b2_water_2} 
Therefore, here we will focus only in the properties of the condensed phases
(liquid and solids).
The ten properties of the test will be the following ones:
\begin{itemize}
\item {1. Vapour-liquid equilibria (VLE) and critical point.}
\item {2. Surface tension.}
\item {3. Densities of the different ice polymorphs.}
\item {4. Phase diagram calculations.}
\item {5. Melting temperature and properties at the melting point.}
\item {6. Maximum in density of water at room pressure (TMD). 
          Values of the thermal expansion coefficient, $\alpha$, and 
          the isothermal compressibility, $\kappa_T$.}
\item {7. Structure of water and ice Ih.}
\item {8. Equation of state at high pressures.}
\item {9. Self-diffusion coefficient.}
\item {10. Dielectric constant.}
\end{itemize}

In order to give an assessment of the quality of the predictions we will 
assign a score to each of properties depending on the predictions of the model.
We recognise that any score is rather arbitrary.
We do not pretend to give an absolute test but rather to give a qualitative
idea of the relative performance of these water models. 
So we have devised a simple scoring scheme: for each property we shall assign 0
points to the model with the poorest performance, 1 to the second, 2 to the 
third and 3 points to the model showing the best performance.

\section {Calculation details}
In this work we compare the performance of TIP3P, TIP4P, TIP5P and TIP4P/2005. 
To perform the comparison we shall use data taken from the literature, either 
from our previous works or from other authors. 
For the cases where no results are available we have carried out new simulations
to determine them. 
In all the calculations of this work the LJ part of the potential has been 
truncated and standard long range corrections are used. Unless stated
otherwise all calculations of this work were obtained by truncating the LJ
part of the potential at 8.5 \AA. 
The importance of an adequate treatment of the long range Coulombic forces when
dealing with water simulations has been pointed out in recent studies.%
\cite{JCP_1998_108_10220,lisal02,rick04JCP_2004_120_06085,yonetani}
In fact, in the case of water, the simple truncation of the Coulombic 
part of the potential causes 
a number of artifacts.\cite{JCP_1998_108_10220,lisal02,rick04JCP_2004_120_06085,yonetani} 
In our view, for water, the simple truncation of the Coulombic part of the potential should be 
avoided and one should use a technique treating adequately the long range 
Coulombic interactions (as for instance Ewald sums or Reaction Field\cite{allen_book,frenkelbook,CPL_1994_231_0366_nolotengo}).
The Ewald sums are especially adequate for phase diagram calculations since 
they can be used not only for the fluid phase but also for the solid phase. 
Because of this, the Ewald summation technique\cite{frenkelbook} has been 
employed in this work for the calculation of the long range electrostatic 
forces.
NpT and phase diagram calculations were done with our Monte Carlo code.
Transport properties were determined by using GROMACS 3.3.\cite{lindahl01} 
Isotropic NpT simulations were used for the liquid phase (and cubic solids) 
while anisotropic Monte Carlo simulations (Parrinello-Rahman like)%
\cite{JAP_1981_52_007182,MP_1985_54_0245} were used for the solid phases.

Recently, we have computed the phase diagram for the TIP4P model by means of
free energy calculations. 
For the solid phases, the Einstein crystal methodology proposed by Frenkel and 
Ladd was used.\cite{frenkel84} 
Further details about these free energy calculations can be found 
elsewhere.\cite{vega_review}
For the fluid phase, the free energy was computed by switching off the charges 
of the water model to arrive to a Lennard Jones model for which the free energy
is well known.\cite{johnson93} 
The free energy calculations for the fluid and solid phases lead to the
determination of a single coexistence point for each coexistence line.
Starting at this coexistence point, the complete coexistence lines were 
obtained by using Gibbs Duhem integration.\cite{kofke93} 
The Gibbs Duhem integration (first proposed by Kofke) is just the numerical 
integration of the Clapeyron equation:
\begin{equation}
\frac{dp}{dT}=\frac{s_{II}-s_I}{v_{II}-v_I}=\frac{h_{II}-h_I}{T(v_{II}-v_I)}
\end{equation}
where we use lower case for thermodynamic properties per particle, and the two 
coexistence phases are labelled as I and II respectively.
Since the difference in enthalpy and volume between two phases can be 
easily determined in computer simulations, the equation can be integrated 
numerically. 
Therefore, a combination of free energy calculations and Gibbs Duhem 
integration allowed us to determine the phase diagram of the TIP4P model.

In this work we have also computed the phase diagram for the TIP3P and TIP5P 
models. 
Instead of using the free energy route to obtain the initial coexistence
point we have used Hamiltonian Gibbs Duhem integration\cite{gdfilosofico1}
which we briefly summarise.
Let us introduce a coupling parameter,$\lambda$, which transforms a potential 
model A into a potential model B (by changing $\lambda$ from zero to one):
\begin{equation}
U(\lambda)=\lambda U_{B} + (1-\lambda)\; U_{A}.
\end{equation}
It is then possible to write a generalized Clapeyron equation as
\begin{equation}
\label{gdftemp}
\frac{dT}{d\lambda}=\frac{T(<u_{B}-u_{A}>_{N,p,T,\lambda}^{II}
-<u_{B}-u_{A}>_{N,p,T,\lambda}^{I})}{h_{II}-h_I}
\end{equation}
where $u_{B}$ is the internal energy per molecule when the interaction between 
particles is described by $U_{B}$ (and a similar definition for $u_{A}$).
If a coexistence point is known for the system with potential A, it is possible
to determine the corresponding coexistence point for the system with potential
B (by integrating the previous equations changing $\lambda$ from zero to one). 
In this way, the task of determining an initial point of the coexistence lines
of the phase diagram of system B is considerably simplified.
The initial coexistence properties for the system A (i.e., the reference system)
must be known.
We have used TIP4P as reference system because its coexistence lines are now 
well known.\cite{sanz1,sanz2}
The phase diagram for TIP3P and TIP5P reported in this work has been obtained
by means of the Hamiltonian Gibbs Duhem integration starting from the known 
coexistence points of the TIP4P model. 
Again, the rest of the coexistence lines have been calculating using the usual
Gibbs-Duhem (Eq. 2) integration.

\section {Results}

{\bf 1. Vapor-liquid equilibria (VLE) and critical point.}

\begin{figure}[!hbt]\centering
\includegraphics[clip,height=7cm,width=0.6\textwidth,angle=-0]{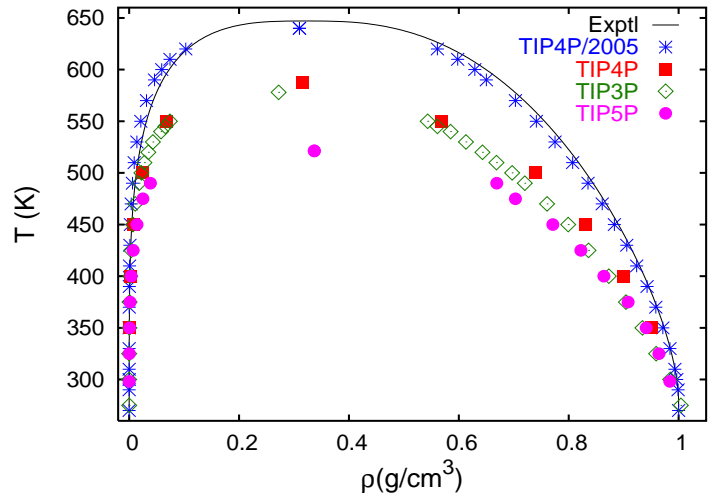}
\caption{Vapor liquid equilibria for TIP3P, TIP4P, TIP5P and TIP4P/2005 models.}
\label{fig_vap_liq}
\end{figure}

The vapour-liquid equilibria of TIP4P and TIP5P has been reported by Nezbeda 
{\em et al.}\cite{lisal01,lisal02,lisal04}
Besides, we have recently calculated the VLE of TIP4P/2005.\cite{vega06} 
However, predictions of the vapour-liquid equilibria for the rigid TIP3P model 
were not found after a literature search (although it was possible to find 
results\cite{tip3p_flexible} for a flexible TIP3P model). 
For this reason we proceeded to evaluate the VLE of the TIP3P model. 
Firstly, we used Hamiltonian Gibbs Duhem integration to obtain an initial 
coexistence point for the TIP3P model at 450~K. 
As initial reference point, we used the coexistence data reported by Lisal 
{\em et al.}\cite{lisal01} of the TIP4P model at 450~K. 
The LJ part of the potential was truncated at a distance slightly smaller than 
half the box length (both for the liquid and for the vapour phase). 
Once the coexistence point at 450~K was obtained for the TIP3P model,
the rest of the coexistence line was obtained from Gibbs-Duhem simulations. 
The coexistence points are presented in Table \ref{tip3p_VLE}. 
The critical properties of all TIP$n$P models are given in Table 
\ref{critical_properties} and Figure~\ref{fig_vap_liq} shows the vapour-liquid 
equilibria for TIP3P, TIP4P, TIP4P/2005 and TIP5P.
 
The lowest critical temperature corresponds to that of the TIP5P model followed
by that of the TIP3P and TIP4P. 
TIP4P/2005 provides a critical point in very good agreement with experiment. 
As it can be seen in Fig. \ref{fig_vap_liq} TIP4P/2005 also provides an 
excellent description of the coexistence densities along the orthobaric curve. 
The description of the critical pressure is not good for any of the models,
which suggest that the inclusion of polarizability is probably required to
reproduce the experimental value of the critical pressure.
Notice that models reproducing the vaporization enthalpy of water without the 
addition of the polarization term (TIP3P, TIP4P and TIP5P) tend to 
predict too low critical temperatures.
The difference between TIP3P and TIP4P is quite small suggesting that three 
charge models yield a critical point around 580~K when forced to reproduce the
vaporization enthalpy. 
This idea is further reinforced by the critical temperature of SPC, which is 
of about 590~K \cite{panagiotopoulos98}.  
The results for TIP5P indicate that the location of the negative charge at the
lone pairs leads does not solve the problem and this model gives the worst
predictions for both the critical temperature and pressure.
The success of TIP4P/2005 in estimating the critical point seems to be related
to the fact that this model does reproduces the vaporization enthalpy only after
the inclusion of the polarization correction (as given by Eq.\ref{polarization}). 
Further evidence of this is obtained from the fact that SPC/E 
(which also incorporates the polarization correction) does also reproduces reasonably 
well the critical temperature of water. 

The correlation between the critical point and the vaporization enthalpy was 
first pointed out by Guillot.\cite{guillot02}
It is obvious that the value of the vaporization enthalpy is related to the 
strength of the hydrogen bond in the model. 
One may try to relate the strength of the hydrogen bond to the magnitude of 
the dipole and quadrupole moments. 
The dipole moments of TIP3P, TIP4P, TIP4P/2005 and TIP5P are rather similar. 
But their quadrupole moments differ significantly. 
TIP5P has the lowest quadrupole moment and TIP4P/2005 has the highest one
and this also true for the critical temperatures. 
However the SPC/E model, which has a relatively low value of the quadrupole 
moment, yields a satisfactory critical temperature. 
Obviously the strength of the hydrogen bond depends not only on the multipole 
moments but also on the parameters of the LJ potential and on the bond geometry. 
From the results of the vapour-liquid equilibria, we assign 3 points to 
TIP4P/2005, 2 points to TIP4P, 1 point to TIP3P and 0 points to TIP5P. 

{\bf 2. Surface tension.}

\begin{figure}[!hbt]\centering
\includegraphics[clip,height=7cm,width=0.6\textwidth,angle=-0]{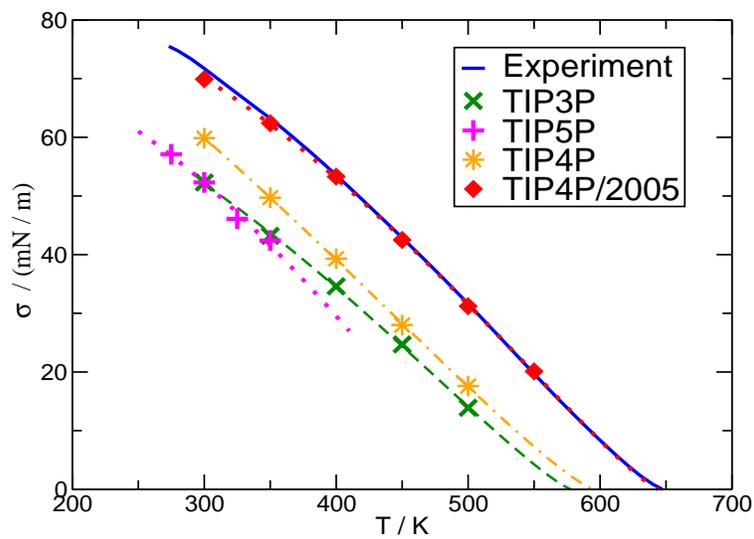}
\caption{Surface tension for different water models.}
\label{fig_surface_tension}
\end{figure}

The values of the surface tension, $\gamma$, for TIP4P and  TIP4P/2005 have been
reported recently.\cite{vega07a} 
For these models the surface tension was obtained using the mechanical 
route\cite{rowlinsonwidom} (from the normal and perpendicular components of 
the pressure tensor) and the test area method\cite{demiguel05} (where 
the Boltzmann factor of a perturbation that changes the interface area but keeps 
constant the total volume is evaluated). 
Results obtained from these two methodologies were in good agreement. 
For TIP5P, results of the surface tension has been reported recently by 
Chen and Smith.\cite{smith_pe_surface_tension} 
In this work we have calculated the surface tension for TIP3P using both the 
mechanical route and the test area method.
Simulations were performed with GROMACS 3.3\cite{lindahl01} for 1024 water 
molecules with an interface area of about ten by ten molecular diameters. 
The length of the runs was 2~ns. 
The LJ part of the potential was truncated at 13 \AA.  
Long range corrections to the surface tension were included as described 
in Ref.~[\onlinecite{vega07a}]. 
The results for TIP3P are presented in Table \ref{st_tip3p}.  
The surface tension of TIP3P can be described quite well by the following 
expression~\cite{internet_equation}:
\begin{equation}
\label{fit}
\gamma = c_{1} \left ( 1- T/T_{c} \right)^{11/9}
 \left[ 1- c_2 \left ( 1- T/T_{c} \right ) \right]
\end{equation}
This expression is used by
the IAPWS (International Association for Properties of Water and Steam)
to describe the experimental values of the surface tension of 
water.\cite{iapws,chaplin}
A fit of the surface tension results for TIP3P furnishes the parameters 
$T_c=578.17$~K, $c_1=176.42~mN/m$, $c_2=0.57857$. 
The critical temperature obtained in this way is in good agreement with that 
obtained from Gibbs Duhem calculations. 
Figure~\ref{fig_surface_tension} shows the surface tension for TIP3P, TIP4P,
TIP4P/2005 and TIP5P.
The lowest values of the surface tension correspond to the TIP5P model followed
by TIP3P and TIP4P.
Again, models reproducing the vaporization enthalpy of water give rather low 
values of the surface tension. 
This is consistent with the lower critical temperature of these models. 
It seems that the TIP4P charge distribution provides a slightly higher value
of the surface tension when compared to TIP3P and TIP5P. 
The predictions of the surface tension of the TIP4P/2005 are in excellent
agreement with experiment\cite{vega07a,franceses_tip4p_2005}. 
How  to explain the performance of the different models? 
Since the surface tension of SPC is similar to that of TIP3P, and that of SPC/E is 
similar to that of TIP4P/2005 (although the predictions of TIP4P/2005 are
even better\cite{vega07a,alejandre08} than those of SPC/E) the use of the polarization correction 
for the vaporization enthalpy seems to be responsible of the improvement of TIP4P/2005
with respect to the TIP3P, TIP4P and TIP5P models. 
Thus, the use of the polarization correction of Berendsen {\em et al.}\cite{berendsen87} 
seems to be 
a prerequisite to obtain models with a good description of the surface tension of water (in
non-polarizable models). 
From the results for the surface tension, we give 3 points to TIP4P/2005, 
2 points to TIP4P, 1  point to TIP3P and 0 points to TIP5P. 
Not surprisingly, the scores of the models are identical to those 
obtained for the vapour-liquid equilibria. 

{\bf 3. Densities of the different ice polymorphs.}

In the book {\em The Physics of Ice},\cite{bookPhysIce} Petrenko and Whitworth
reported the density for several solid phases of water at a certain 
thermodynamic state. These states are used often to test the performance
of water models\cite{sanz1,abascal05b,baranyai05}.
Simulation results\cite{sanz1,pccpgdr,abascal05a,abascal05b} indicate that
TIP5P overestimates the density of the ices by about seven per cent whereas
TIP4P overestimates the densities by less than three per cent and TIP4P/2005 
gives an error of less than one per cent.
The failure of TIP5P is probably a consequence of a too short distance between
the oxygens when the hydrogen bond is formed.\cite{wales_water_clusters}
This might also be related to the fact that the negative charge is too far away
from the oxygens in TIP5P.
In this work we have obtained the predictions for the TIP3P and SPC models
which have not been reported yet. 
We have carried out NpT simulations with anisotropic scaling. 
For the initial configurations we used the structural data obtained from 
diffraction experiments. 
This is all what is needed for ices in which the protons are ordered (II, IX, 
XI, VIII, XIII, XIV). 
For ices with proton disorder the oxygens were located using crystallographic
information and a proton disordered configuration (with no net dipole moment 
and satisfying the Bernal-Fowler rules\cite{bernal33,pauling35,macdowell04}) was generated with the algorithm
proposed by Buch {\em et al.}\cite{buch98}.  
Notice that, although we use experimental information to generate the initial 
solid configuration, the system can relax since we are using the anisotropic NpT 
scaling in the simulations.

The ice densities for the TIP3P and SPC models are given in Table 
\ref{density_ices}.  
SPC yields errors of about 2.2\%.
In general all solid phases were mechanically stable with the SPC model, the 
only exception being ice XII that melted spontaneously at the simulated 
temperature. 
The situation is much worse for the TIP3P model for which several of the 
solid phases melted spontaneously (in particular ice Ih, III, V and XII).
As it will be discussed later this is related to the extremely 
low melting points of solid phases for the TIP3P model.\cite{vegafilosofico}
Table~\ref{density_ices} also reports the internal energies for the different
crystalline phases of TIP3P and SPC (those for other models can be found
elsewhere\cite{abascal05b,aragones1,aragones2}).

\begin{figure}[!hbt]\centering
\includegraphics[clip,height=15cm,width=0.6\textwidth,angle=-90]{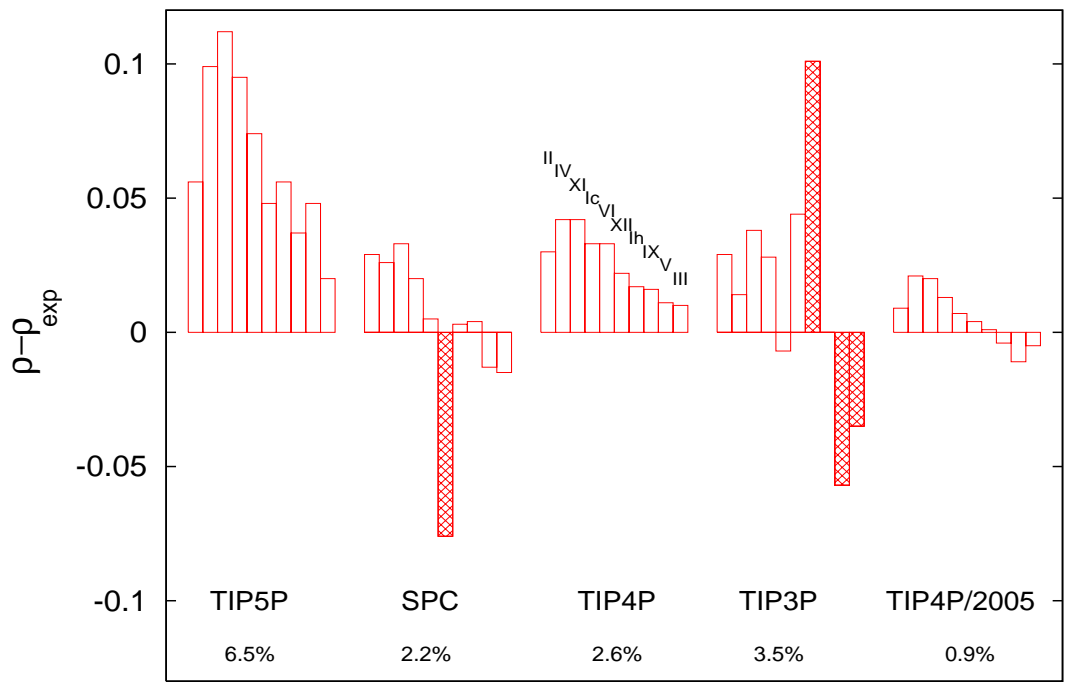}
\caption{Deviations in the density predictions for several ice polymorphs at
the thermodynamic states reported by Petrenko and Whitworth.\cite{bookPhysIce}
The experimental value of the density for ice II is taken from Fortes {\em et 
al.}\cite{fortes_abinitio,fortes_jac05} (instead of the value of Kamb {\em et 
al.}\cite{kamb_iceII,JCP_1971_55_01934_nolotengo} used in our previous work). 
For TIP3P and SPC, the filled rectangles indicates that the corresponding solid
phase melts and that the density of the liquid was used to compute the error.}
\label{fig_ices_densities}
\end{figure}

Figure~\ref{fig_ices_densities} represents the average deviation from experiment
for several ice polymorphs.
For the experimental density of ice II we are using the value recently
reported by Fortes {\em et al.},\cite{fortes_jac05} which evidenced that the
value of Kamb\cite{kamb_iceII} could be distorted by the presence
of helium inside the ice II structure (see also the discussion in Ref.
[\onlinecite{aragones2}]).
In fact, the simulation results agree much better with the new reported data of
Fortes {\em et al.} reinforcing the idea that the value reported by Kamb
is probably incorrect.

According to the results of Fig.~\ref{fig_ices_densities}, we assign 3 points
to the performance of TIP4P/2005, 2 points to TIP4P, 1 point to TIP5P and
0 points to TIP3P (the low score of TIP3P is due to the fact that several ices
melt well below their experimental melting temperatures so that this model is
not useful to study the behaviour of the solid phases of water).

{\bf  4. Phase diagram calculations.}

\begin{figure}[!hbt]\centering
\includegraphics[clip,height=6cm,width=0.47\textwidth,angle=-0]{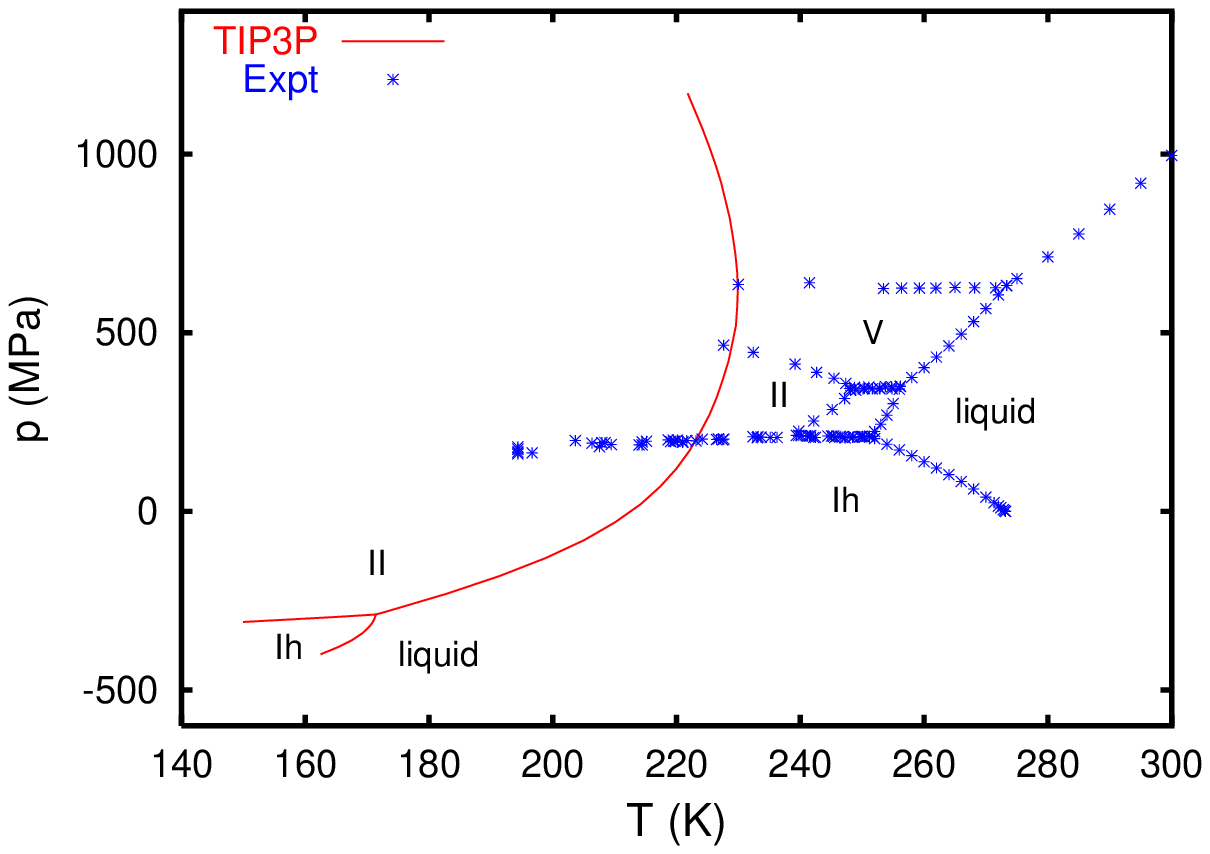}
\includegraphics[clip,height=6cm,width=0.47\textwidth,angle=-0]{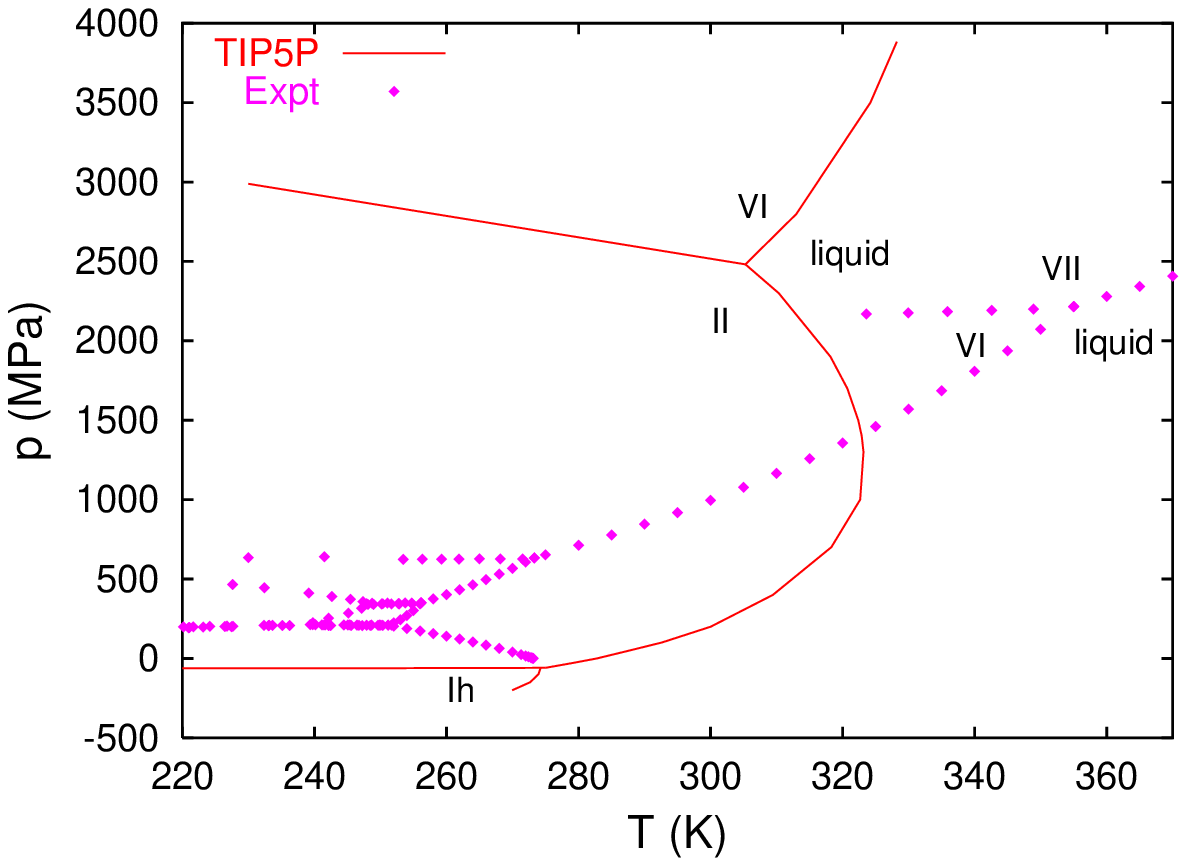}
\caption{Phase diagram as obtained in this work for TIP3P and TIP5P. 
Symbols: Experimental phase diagram. Lines: Computed phase diagram.
Left. Results for the TIP3P model. Right. Results for the TIP5P model.}
\label{fig_phase_diagram_tip3p_tip5p}
\end{figure}

\begin{figure}[!hbt]\centering
\includegraphics[clip,height=7cm,width=0.6\textwidth,angle=-0]{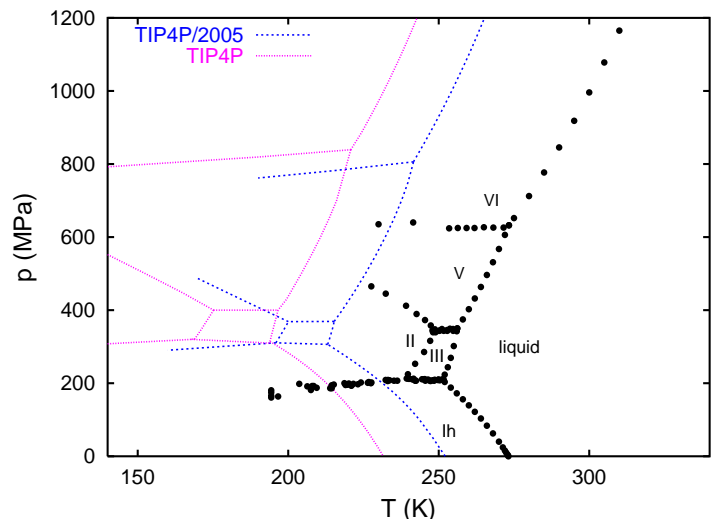}
\caption{Phase diagram of TIP4P and TIP4P/2005 models. 
Symbols: Experimental phase 
diagram, lines: computed phase diagram.}
\label{fig_phase_diagram_tip4p_tip4p_2005}
\end{figure}

The phase diagram of TIP4P, SPC/E and TIP4P/2005 has been determined in previous work.
\cite{sanz1,sanz2,abascal05b,vegautrecht}
The phase diagram for TIP3P has not been determined so far.
In this work we have used Hamiltonian Gibbs-Duhem integration to compute it
and to complete the previous calculations for TIP5P.
The results are presented in Figure~\ref{fig_phase_diagram_tip3p_tip5p}.
The predictions of TIP3P and TIP5P are quite poor.
Ice Ih is thermodynamically stable only at negative pressures.
The stable phase at the normal melting point for both models is ice II.
This surprising result has been confirmed recently by evaluating the properties
of ices at zero temperature and pressure.\cite{aragones1}
The performance of TIP4P/2005 is quite good (Figure
\ref{fig_phase_diagram_tip4p_tip4p_2005}).
The overall performance of TIP4P is also quite good although somewhat shifted
to lower temperatures as compared to TIP4P/2005.

What is the origin of the success of TIP4P/2005 and TIP4P?
Certainly, it can not be related to the the value of the vaporization enthalpy.
This is clear since TIP4P/2005 and TIP4P give good phase diagram predictions
and the first use the polarization correction whereas the second does not.
Since the bond geometry of all models is the same, the different performance
between them should be related to the differences in the charge distribution.
It is not surprising that the solid phases provide information about the
orientational dependence of the pair potential in the molecule of water.
After all, in the solid phase the molecules adopt certain relative
orientations which define the structure of the crystal.
Polar forces are strongly dependent on the orientation.
Since the dipole moment is similar for TIP3P, TIP4P, TIP4P/2005 and TIP5P, the
different predictions found in Figs.~\ref{fig_phase_diagram_tip3p_tip5p} and
\ref{fig_phase_diagram_tip4p_tip4p_2005} must be related to differences
in higher multipolar moments.
In fact, as it was discussed above, the quadrupole moments for these water
models are quite different.
We have recently reported that the ratio of the dipole to quadrupole moment
seems to play a crucial role in the quality of the phase diagram predicted
by the different water models.\cite{abascal07b}
As it can be seen in Table \ref{tabl_moments}, the ratio $\mu/Q_T$ for TIP4P
and TIP4P/2005 is about one whereas for TIP3P it increases to 1.33.
For the TIP5P model the ratio is even larger.
In summary, a qualitatively good description of the phase diagram of water
requires a reasonable balance between dipolar and quadrupolar forces, and the
factor affecting this ratio is just the charge distribution.
Not surprisingly, the phase diagram prediction for SPC/E (with a charge
distribution similar to that of TIP3P) is also quite poor.
Therefore, we assign 3 points to TIP4P/2005, 2 points to TIP4P, 1 point to
TIP5P and 0 points to TIP3P.

{\bf 5. Melting temperature. Properties at melting.}

Obtaining the melting point of water models is not as obvious as it may appear.
The simplest approach (which works fine in the lab\cite{bridgman12})
of heating the ice within the simulation box and determining the temperature at which melts, does not provides the true melting point.
In fact, when NpT simulations are performed under periodical boundary
conditions, ices usually melt at a temperature about 80-90~K above the true
melting point\cite{mcbride1} ({\em i.e.}, the temperature at which the chemical
potential of the liquid and solid phases are identical).
But, in real experiments, ices can not be superheated (at least for
a reasonable time).
The absence of a free surface is responsible for the superheating of ices in
NpT simulations.
In fact, when the simulations are performed with a free surface, super heating
is suppressed.\cite{maria06}
For this reason the evaluation the melting point of water models requires 
special techniques.
The melting point of TIP4P was obtained from free energy calculations of both
the fluid and the solid phase.
Details of these free energy calculations have been given 
recently.\cite{vega_review}  
The melting temperatures of TIP5P, TIP3P and TIP4P/2005 were obtained with
Hamiltonian Gibbs-Duhem integration using TIP4P as a reference model.
The melting points obtained via the free energy route are in complete agreement
with those obtained from direct coexistence 
simulations\cite{haymettip4pmelting,ramon06,wang05,quigley08}
where the fluid and solid phases are kept in contact within the simulation box.

We now discuss the melting temperatures of ice Ih for TIP3P, TIP4P, TIP5P, 
TIP4P/2005 at the standard pressure of 1 bar.
It should be mentioned once again that for TIP3P and TIP5P, ice II is more
stable than ice Ih at ambient pressure.
It is possible to locate the melting point of ice Ih, even though is metastable,
provided that it is mechanically stable.
For TIP5P the melting point of ice II is about 2~K above that of ice Ih but
the melting point of ice II for TIP3P is about 60~K above that of ice Ih.
In Table \ref{propiedades_melting}, the melting temperature of ice Ih (at 1
bar) and the properties of the solid and the liquid phases at coexistence are
given for these models.
Concerning the melting temperature, it is as low as 146~K for TIP3P.
Thus, TIP3P is probably the poorest model to study solid phases of water.
Since it is possible to simulate ices up to temperatures about 80~K above the
melting point, the temperature of 230~K is roughly speaking the highest one
that can be used to study ice Ih with the TIP3P model.
At higher temperatures the TIP3P ice Ih melts.
Our estimates for the melting temperature of TIP4P, TIP4P/2005 and TIP5P are
approximately 232~K, 252~K, and 274~K, respectively.
The result for TIP4P is in good agreement with the value reported by Gao 
{\em et al}.\cite{JCP_2000_112_08534} and Koyama {\em et al}\cite{tanaka04}.
Another interesting quantity is the ratio of the normal melting temperature
to the critical temperature, $T_m/T_c$.  
Since the difference between the normal melting temperature and the triple 
point temperature is of about 0.01~K, this ratio determines the range of 
existence of the liquid phase for the considered model. 
Experimentally, $T_m/T_c$=0.42. 
The value of the ratio for TIP4P models (TIP4P and TIP4P/2005) is essentially
the same, 0.394, but it increases to 0.525 for TIP5P.
Thus TIP4P models describe significantly better the experimental range of 
existence of the liquid phase for water. 
For TIP3P the ratio is considerably lower, 0.25, when the melting temperature 
of ice Ih (146~K) is considered.
But it increases to 0.36 using 210~K as the melting temperature for the stable 
phase at normal melting for the TIP3P model which is ice II.

Table \ref{propiedades_melting} also presents the coexistence properties
(densities at coexistence, slope of the melting curve $dp/dT$ and enthalpy of
melting).
TIP4P/2005 provides the best estimates of the coexistence densities and
TIP5P gives a poor estimate of the density of ice Ih, and a completely wrong
prediction of the slope $dp/dT$.
This is because, for TIP5P, the densities of ice Ih and of the liquid are quite
similar.
On the other hand, no model is able to reproduce the melting enthalpy.
It is too small (three times lower) for TIP3P. The  
TIP4P and TIP4P/2005 models do also underestimate the melting enthalpy but the errors
are much smaller, about 30\% and 20\%, respectively.
Finally, TIP5P overestimates the melting enthalpy by approximately a 20\%.

Let us now try to provide a rational basis for the previous results.
For three charge models (TIP3P, TIP4P and TIP4P/2005) there is a clear 
correlation between the melting temperature of ice Ih and the quadrupole 
moment of the molecule.\cite{abascal07a,abascal07c} 
Our conclusion is that models locating the negative charge on the oxygen atom 
have low melting point temperatures for ice Ih. 
Locating the negative charge shifted from the oxygen along the H-O-H bisector 
increases the melting temperature. 
The improvement is better if the polarization correction is used in the 
calculation of the vaporization enthalpy as a target property. 
When this is done (as in TIP4P/2005) the melting point is about 20~K below the 
experimental value.
This is the closest you can get for the melting point of ice Ih with a three 
charge model while still describing the vaporization enthalpy with the polarization 
correction of Berendsen {\em et al.}\cite{berendsen87}. 
The only way of reproducing the experimental melting point of ice Ih within three 
charge models is to sacrify the vaporization enthalpy as a target property. 
In fact, we have developed a model (denoted as TIP4P/Ice)\cite{abascal05a} 
which reproduces the melting point of ice Ih but overestimates the vaporization
enthalpy.
The behaviour of TIP5P is different, it yields a good prediction for the
melting temperature in spite of having a low quadrupole moment. 
Probably the fact that there are two negative charges instead of just one
is the responsible of this different behaviour.
It is likely that locating the negative charge on the lone pair electrons 
enhances the formation of hydrogen bonds in the solid phases provoking an 
increase in the melting point. 

TIP5P reproduces the melting temperature of water.
But this is not for free, since the estimate of the coexistence density of ice 
and of the dp/dT slope becomes then quite poor.
Conversely, TIP4P/2005 predicts a rather low melting temperature but it
gives quite reasonable estimates for the properties at melting.
For this reason, there is no clear winner in this test so we have decided to 
assign $2.5$ points to both TIP5P and TIP4P/2005, 1 point to TIP4P and 0 points
to TIP3P. 

{\bf  6. Temperature of maximum density. Thermal coefficients $\alpha$
and $\kappa_T$.}

\begin{figure}[!hbt]\centering
\includegraphics[clip,height=7cm,width=0.6\textwidth,angle=-0]{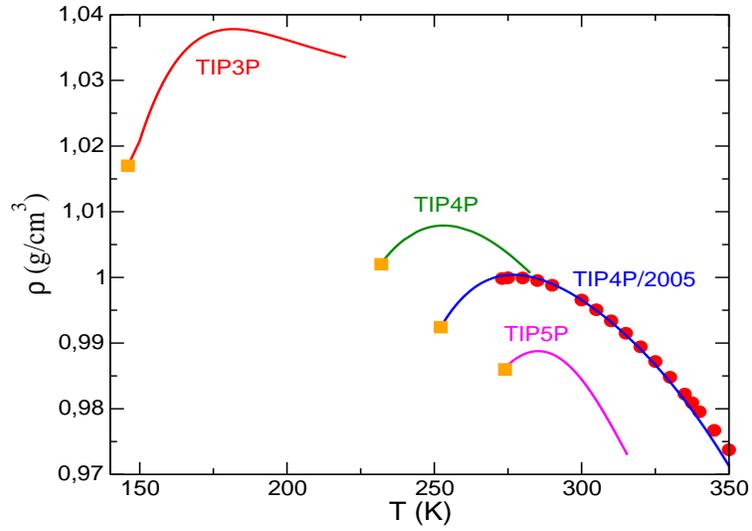}
\caption{Maximum in density for several water models at room pressure. 
Filled circles: experimental results. Lines: simulation results. 
For each model the square represents the location of the melting temperature of 
the model for ice Ih.}
\label{fig_tmd}
\end{figure}

The density of liquid water for the room pressure isobar as a function of the 
temperature for TIP3P, SPC and TIP4P with a simple truncation of the Coulombic 
forces have been reported by Jorgensen and Jenson.\cite{jorgensenjcc}. 
Recently, we have also calculated the liquid densities at normal 
pressure\cite{vega05b,abascal05b} for TIP3P, TIP4P, TIP5P and TIP4P/2005 using
Ewald sums to deal with long range Coulombic forces. 
Good agreement between our results and those reported by Paschek was
found.\cite{paschek} 
Our results are shown in Figure~\ref{fig_tmd}. 
All models do exhibit a maximum in the density of water along the isobar. 
For some time it was believed that TIP3P did not exhibit this maximum but 
it is now clear that the maximum also exists for this model (although located 
at very low temperatures).  
In Table \ref{tmd_alpha_kappa}, the location of the TMD and the values of the 
thermal expansion coefficient ($\alpha$ ) and of the isothermal compressibility
($\kappa_T$) at room temperature and pressure are presented. 
It is clear that the poorest description of the TMD is provided by TIP3P. 
The behaviour of TIP4P is noticeably better. 
TIP4P/2005 reproduces nicely the density of water (and its maximum) for all
the temperatures along the room pressure isobar. 
We have shown recently\cite{vega05b} that, ---for three charge models as TIP3P, 
TIP4P and TIP4P/2005--- the difference between the TMD and the melting 
temperature of ice is around 25~K (the experimental difference is 4~K). 
For this reason, the location of the TMD correlates very well with the 
melting temperature. 
TIP5P also reproduces the temperature at which the maximum density in water 
occurs.
For the TIP5P the difference between the melting point of ice Ih and the 
temperature of the maximum in density is about 11~K.  
In Fig.\ref{fig_tmd} the densities of TIP5P as obtained from simulations using 
Ewald summation are presented. 
Notice that TIP5P does not reproduce the experimental densities when Ewald 
summation is used.  
In the original paper where the TIP5P model was proposed by Mahoney and
Jorgensen the potential was truncated at 9 \AA. 
In these conditions TIP5P reproduces the density of water at the maximum.
It has been noticed by other authors that in general, lower densities are 
obtained when Ewald sums are implemented than when the electrostatic potential
is merely truncated at a given distance. 
The decrease in density is particularly noticeable in the case of the TIP5P 
model. 
For this reason the maximum in density of the TIP5P model occurs at 285~K when 
Ewald sums are used\cite{vega05b,lisal02} whereas the maximum takes place at 
277~K when the potential is truncated at 9 \AA.  
Fig.\ref{fig_tmd} shows that for the TIP5P model 
when the whole normal pressure isobar is 
considered (not just the maxima), the curvature is not correct. 
In other words, the dependence of the density with temperature at constant
pressure (given by the thermal expansion coefficient $\alpha$) is not properly 
predicted by TIP5P (see Table \ref{tmd_alpha_kappa}).
Importantly, this is also true when the potential is truncated at 9 \AA . 
Therefore, TIP5P does not reproduces correctly $\alpha$, neither with 
the potential truncated at 9 ~\AA\ nor with Ewald sums.  
In summary, both TIP4P/2005 and TIP5P reproduces the temperature at which the 
density maximum occurs. 
This is by design since the TMD was used as a target property in both models. 
However, the description of the complete room pressure isobar is much better in 
the TIP4P/2005 which reproduces nicely the whole curve providing a very good 
estimate of the coefficient $\alpha$. 
Also the estimate of $\kappa_T$ is better for TIP4P/2005. 
The excellent description of the densities along the isobar has an interesting 
consequence. 
Since the vapour pressure of water is quite small up to relatively high 
temperatures, orthobaric densities are essentially identical to those obtained 
from the room pressure isobar. 
Therefore, a model describing correctly the equation of state along the room 
pressure isobar will also provide reliable estimates of the orthobaric densities(at least, at temperatures not too close to the critical one). 
Thus the good description of the ambient pressure isobar of TIP4P/2005 explains
in part the good description of the coexistence curve presented previously.  
Although TIP5P estimates correctly the location of the TMD (when using a cutoff
for the electrostatic interactions), it does not yield satisfactory densities
for the the rest of the the room pressure isobar and this explains the poor 
performance in describing orthobaric coexistence densities. 
Thus, using the TMD as a target property is a good idea, but it is far better 
to use the complete room pressure isobar (of course including the maximum) as 
a target property. 
For future developments of water potentials we do really recommend to use the 
complete room pressure isobar as a target property.
According to the discussion above, we give 3 points to TIP4P/2005, 2 
points to TIP5P, 1 point to TIP4P and 0 points to TIP3P. 

{\bf 7. Structure of liquid water and ice.}

\begin{figure}[!hbt]
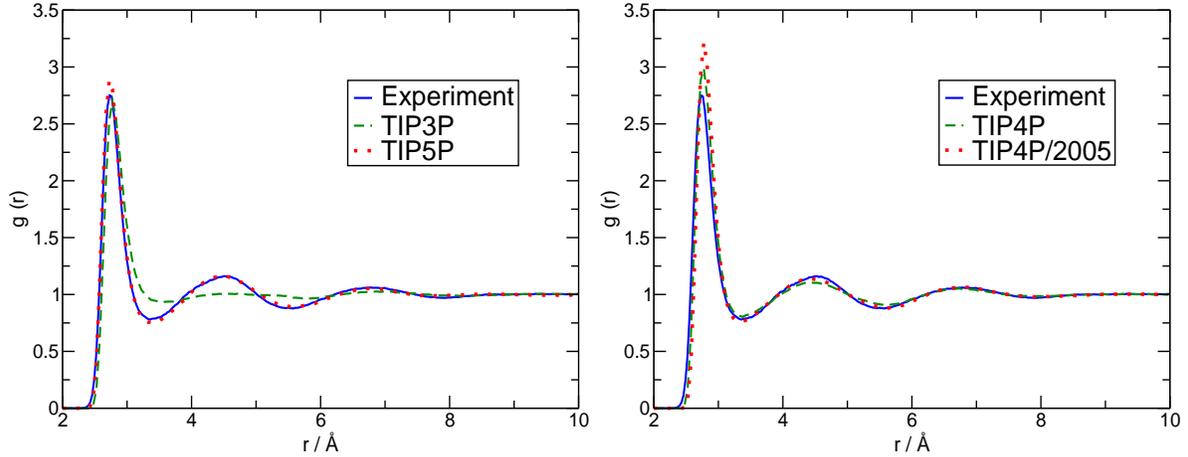
\centering
\includegraphics[clip,height=6cm,width=0.47\textwidth,angle=-0]{fig7a.eps}
\includegraphics[clip,height=6cm,width=0.47\textwidth,angle=-0]{fig7b.eps}
\caption{ Oxygen-oxygen radial distribution function for liquid water  
at $T=298$~K and 
$p=1bar$. Experimental results were taken from Soper\cite{soper00}. 
Left, results for models TIP3P and TIP5P. 
Right, results for TIP4P and TIP4P/2005.}
\label{fig_rdf_water}
\end{figure}

\begin{figure}[!hbt]
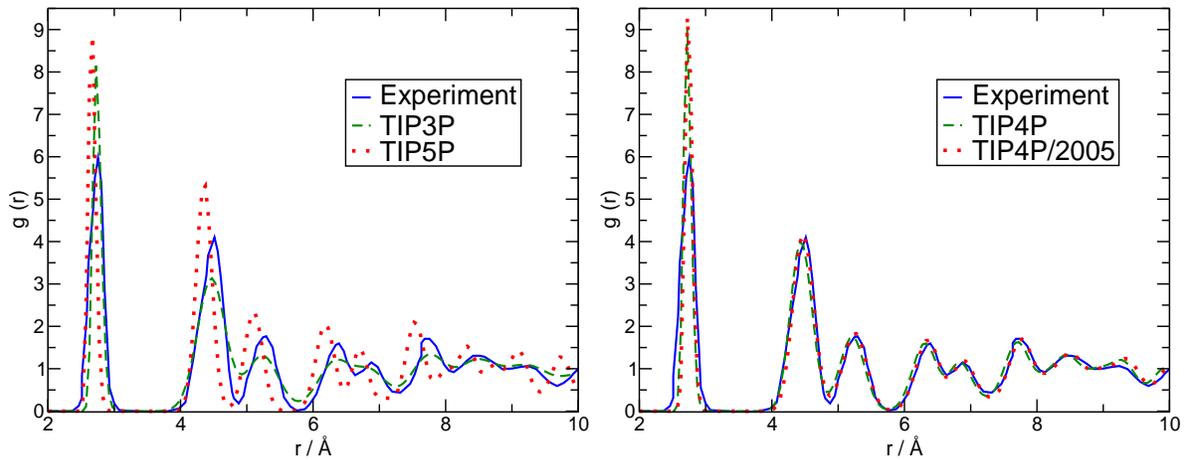
\centering
\includegraphics[clip,height=6cm,width=0.47\textwidth,angle=-0]{fig8a.eps}
\includegraphics[clip,height=6cm,width=0.47\textwidth,angle=-0]{fig8b.eps}
\caption{
Oxygen-oxygen radial distribution function for ice Ih at
$T=77$~K and $p=1bar$.
Experimental results as reported by 
Narten {\em et al.} \cite{JCP_1976_64_01106}
Left, results for models TIP3P and TIP5P . Right , results for 
TIP4P, TIP4P/2005.}
\label{fig_rdf_ice_Ih}
\end{figure}

The experimental oxygen-oxygen radial distribution function for liquid water 
and for ice Ih will be compared to the predictions from the simulations. 
In Figure~\ref{fig_rdf_water} the comparison is made for liquid water. 
Experimental results are taken from Soper.\cite{soper00}
TIP5P provides the best estimate of the radial distribution function. 
The predictions of TIP4P/2005 are good but the first peak is too high. 
TIP4P gives quite acceptable predictions but not as good as the previous models.
Once again, the discrepancy between the results for TIP3P and experiment are
notorious.
In figure~\ref{fig_rdf_ice_Ih} the comparison is made for ice Ih at 77~K and 
1~bar. 
Experimental results are taken from Narten {\em et al.}\cite{JCP_1976_64_01106} 
The predictions of TIP4P/2005 are now the best (although it overestimates again
the height of the first peak). 
The results for TIP4P are also very good.
Both TIP5P and TIP3P fail in the description of the structure of the solid
beyond the first coordination shell. 
TIP4P/2005 provides good estimates of the radial distribution functions but it 
overestimates the height of the first peak. 
It is likely than quantum effects (which could be determined by using path 
integral simulations) may be significant to understand the amplitude of the 
first maximum.
Indeed, this has been shown to be the case not only for liquid 
water\cite{berne01,kusalik04} but also for ice Ih as reported by Kusalik and 
de la Pe\~na\cite{kusalik_jacs05,kusalik_jcp05}. 
Further work on this point is probably needed. 
There is no clear winner concerning structural predictions (TIP5P performs 
better for water and TIP4P/2005 for ice Ih). 
For this reason we have decided to give $2.5$ points to both TIP5P and 
TIP4P/2005, one point to TIP4P and zero points to TIP3P. 

{\bf 8. Equation of state of liquid water at high pressures.}

\begin{figure}[!hbt]\centering
\includegraphics[clip,width=0.8\textwidth,angle=0]{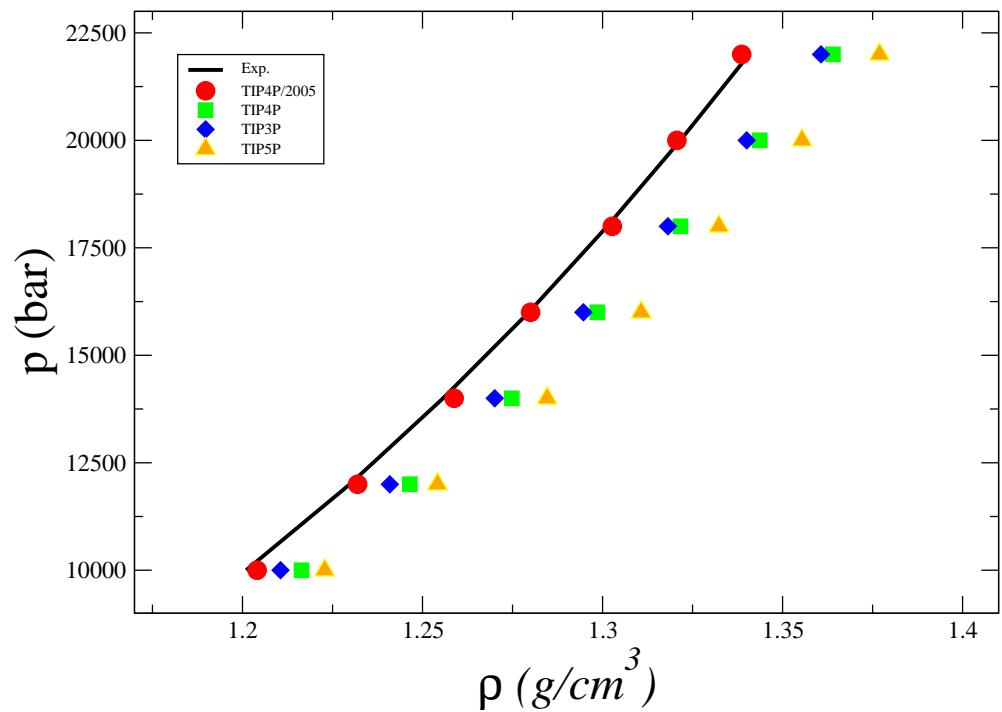}
\caption{Equation of state at $T=373$~K for several water models.}
\label{eos_373}
\end{figure}

In a number of geological applications the equation of state (EOS) of water 
at high temperatures and pressures is needed. 
Therefore it is of interest to analyse the capacity of these models
to predict the behaviour of liquid water at high pressure.
Figure ~\ref{eos_373} displays the EOS for TIP3P, TIP4P, TIP4P/2005 and TIP5P 
along the 373~K isotherm for pressures up to 24000bar (obtained with 360 water
molecules). 
Experimental results are taken from the EOS of Wagner\cite{JPCRD_1994_23_0515} 
and from the experimental measurements of Brown {\em et al.}\cite{brown04}. 
TIP4P/2005 provides an excellent description of the EOS.
The differences with experimental data increase for TIP3P followed by TIP4P. 
The performance of TIP5P is quite poor.
We thus assign 3 points to TIP4P/2005, 2 points to TIP3P, 1  point to 
TIP4P and 0 points to TIP5P. 

\begin{figure}[!hbt]\centering
\includegraphics[clip,scale=0.4,angle=-90]{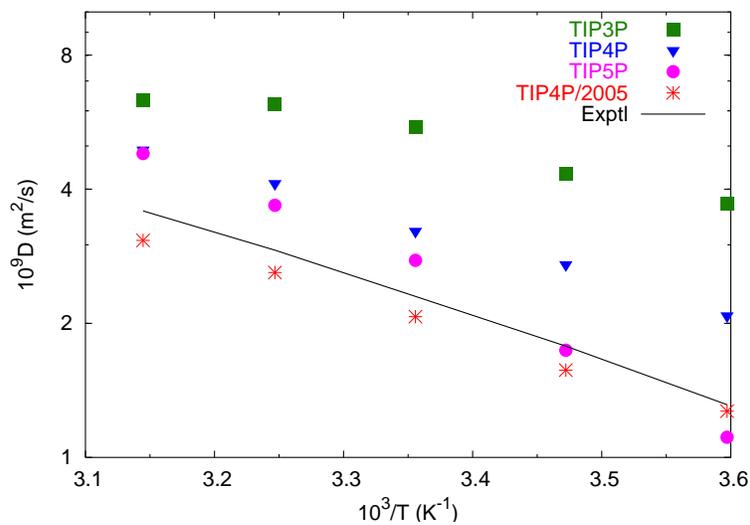}
\caption{Logarithmic plot of the self-diffusion coefficients versus the inverse
of the temperature}
\label{fig_d}
\end{figure}

{\bf 9. Self-diffusion coefficient. }
We have computed the self-diffusion coefficient as a function of temperature 
(at 1 bar) for TIP3P, TIP5P, TIP4P and TIP4P/2005. 
Simulations were performed with the GROMACS\cite{lindahl01} package and the 
diffusion coefficients were determined from the slope of the mean square 
displacement versus time (using 360 molecules). 
A relaxation time of 5~ps was used  for the thermostat and for the barostat.
Results are presented in Table \ref{tablaD}. 
The diffusion coefficient of TIP3P is too high at all the temperatures
investigated.
This suggests that the hydrogen bonding for this model is probably too weak.
Likely, the weakness of the hydrogen bond is also responsible of the low 
ice Ih melting temperature and the loss of structure for the liquid phase
beyond the first coordination shell.
This may be a big concern in simulation studies of proteins where there must 
be a competition between intramolecular and intermolecular hydrogen bonds. 
The diffusion coefficients of TIP4P are closer to experiment but they are
still too high reflecting the fact that using the vaporization enthalpy 
as a target property leads to highly diffusive water models.
TIP5P yields results in agreement with experimental data in the vicinity of 
300 K but the departures from experiment greatly increase as the temperature 
moves away from the ambient one.
In fact, at 318 K the result furnished by TIP5P is almost the same as that
for TIP4P.
Figure~\ref{fig_d} shows that the slope of the line log D {\em vs} 1/T is quite 
poor (TIP4P and even TIP3P are superior in this aspect).
This seems to reflect the overall results of TIP5P: good results at ambient
conditions but becoming increasingly bad as one moves away from that point.
The best predictions are those of TIP4P/2005.
Although a little bit below the experimental values they show the correct trend
in its dependence with temperature. 
Within three charge models, the use of the polarization correction  
to describe the 
vaporization enthalpy leads to water models with improved diffusion properties. 
That was already true for SPC/E (which provides much better diffusion 
coefficients than SPC) and seems also to be true for TIP4P/2005 (which provides
much better estimates of the diffusion coefficients than TIP4P).
We give then 3 points to both TIP4P/2005, 2 points to TIP5P, 1 point to 
TIP4P and zero points to TIP3P. 

{\bf 10. Dielectric constant. }

Let us finish by presenting results for the dielectric constant of liquid 
water at room temperature and pressure. 
Again, simulations were performed with the package GROMACS\cite{lindahl01} 
at room temperature and pressure. 
The simulations lasted 8~ns, and the system consisted of 360 molecules.
The value of the dielectric constant is presented in Table \ref{que_pasa}. 
TIP5P provides the best estimate of the dielectric constant followed by TIP3P. 
TIP4P yields the worse value for the dielectric constant.
TIP4P/2005 predicts a better dielectric constant than TIP4P but it is still 
far from the experimental value.
It is obvious that the TIP4P charge distribution tends to give low dielectric 
constants. 
This is the only property for which the TIP4P charge distribution is in 
trouble against charge distributions with the negative charge located at the 
oxygen atom.
In fact, the dielectric constants of SPC and SPC/E are better than those of 
TIP4P and TIP4P/2005, respectively, indicating that locating the charge on the 
oxygen tends to give better predictions for the dielectric constant. 
Recently, Rick\cite{rick04JCP_2004_120_06085} has studied in detail the 
behaviour of the dielectric constant as a function of the dipole and quadrupole
moments for different water models. 
He has shown that a larger dipole increases the value of the dielectric 
constant whereas a larger quadrupole decreases it.
He proposed an equation to correlate the dielectric constant of water models
with these multipole moments:
\begin{equation}
   \epsilon = -85 + 98  \mu - 35.7 Q_{T}
\end{equation}
where the dipole moment is given in Debye and $Q_{T}$ in (Debye \AA).
The dipole moments of all water models are quite similar. 
However they differ significantly in the value of the quadrupole moment. 
Thus, the lower value of the dielectric constant for TIP4P models is a direct 
consequence of the higher quadrupole moment of these type of models.
The higher quadrupole moment of TIP4P/2005 was quite good to improve predictions
for melting point and phase diagram. 
Unfortunately, it also seems responsible for the low dielectric constant of the
model. 

The dielectric constant is given by the fluctuations of the dipole moment of 
the sample. 
It is thus related to the instantaneous values of the polarization of the 
sample. 
Non-polarizable models attempt to incorporate (in a mean field way) the effect 
of the polarizability. 
Thus, the dipole moments of non-polarizable models are higher than those of 
the gas phase. 
It is likely that a property as the dielectric constant, depending so 
dramatically on just the dipole moment fluctuations, can be hardly reproduced 
by an effective potential in which the molecular charge cannot fluctuate.
In fact, polarizable models tend to have higher dielectric 
constants\cite{cummings05} than their non-polarizable counterparts. 
If this is the case, the good agreement of TIP3P and TIP5P may be somewhat 
forced.
Probably, polarizable versions of TIP3P and TIP5P will tend to give 
too high dielectric constants whereas the introduction of polarizability will 
improve the predictions of models based on the TIP4P charge distribution.
Further work is required to analyse this issue in more detail since it is 
difficult at this stage to establish definitive conclusions. 
Concerning the dielectric constant predictions, we give 3 points to TIP5P, 
2 points to TIP3P, 1 point to TIP4P/2005 and 0 points to TIP4P.   

\section {Conclusions}
Table \ref{final_exam} summarizes the scores obtained by the models for each 
of the test properties.
The final result is that TIP4P/2005 has obtained 27 points, TIP5P 14, TIP4P 13 
and TIP3P 6 over a maximum possible number of 30 points.   
For most of the properties TIP4P/2005 yielded the best performance. 
The main exception is the dielectric constant for which TIP4P/2005 yields a too low value. 
For the second position, TIP5P and TIP4P obtained very similar scores.
TIP5P improves the melting point predictions, TMD, dielectric constant and 
diffusivity with respect to TIP4P, but it is clearly worse in phase diagram 
predictions, critical point, density of ices, and high pressure behaviour. 
In that respect TIP5P and TIP4P yielded a similar performance and the choice 
for the one or the other potential may depend on the considered property to 
study.  
The less satisfactory model, well below any of the others is TIP3P (see Table \ref{final_exam}).
Somewhat surprisingly TIP3P is probably the most used model in simulations of biomolecules. 
In our opinion the only reason to continue using TIP3P is that certain force fields were 
optimized to be used with TIP3P water. 
It is not fully obvious whether the force fields must be used with a given
water model. 
Some researchers have challenged this idea.\cite{force_field_and_water_model} 
In any case, it is clear that new force fields should also be built  around 
better water models.

We have not included in the comparison SPC or SPC/E models. 
The performance of SPC is certainly better than that of TIP3P. 
This is due to the fact, that if the negative charge is located on the oxygen, 
the larger OH bond length of SPC, and the tetrahedral bond angle increases
the values of the quadrupole moment and this improves the performance of the 
model. 
However, SPC/E yields an overall better performance than SPC.  
In fact, it improves the prediction of almost all of the properties. 
The performance of SPC/E, in case it would have been included in the test, 
would have been better than that of TIP5P and TIP4P but well below 
that of TIP4P/2005.
The number of points obtained by SPC/E would have been about 21 points 
(3 for vapour-liquid equilibria, 2 points for surface tension, 2 points for 
the density of ices, 1 point for the phase diagram, 1 point for the melting 
properties, 1 point for the TMD, $\alpha$ and $\kappa_T$, 2 points for structure
predictions, 3 points for the equation of state at high pressures, 3 points for
the diffusion coefficient and 3 points for the dielectric constant). 
Thus, overall SPC/E improves the predictions with respect to TIP4P and TIP5P 
but it is well below the number of points obtained by TIP4P/2005.
 

Why the performance of TIP4P/2005 is in general so good?
TIP4P/2005 is just a TIP4P model with the hydrogen charge augmented from 
0.52~$e$ to 0.556~$e$ and with the value of $\epsilon/k_{B}$ increased from 
73~K to 93~K (the values of $\sigma$ and d$_{OM}$ are quite similar in both 
models). 
It has taken more than twenty years to realize that these two small changes 
improve dramatically the performance of the model. 
The steps  leading to TIP4P/2005 are basically three.  
First, the realization of the fact that TIP4P provided a better phase diagram 
than SPC/E.
Secondly, the attempt to improve the prediction of the melting temperature
could not be done without accepting the idea of introducing a correction term 
in the calculation of the vaporization enthalpy.
The idea, first introduced by Berendsen\cite{berendsen87} {\em et al.}, enabled
them to transform SPC into SPC/E which, in our opinion, is a much better model 
than SPC. 
The same could also be useful within the TIP4P geometry, and the final result 
is TIP4P/2005. 
But there was an additional refinement.
Using the TMD  as a target property (as first done by TIP5P) or, even better,
fitting the whole room pressure isobar could improve the overall performance. 
Thus, TIP4P/2005 incorporated this property as a target property. 
In summary, TIP4P/2005 takes from TIP4P the charge distribution (which yields 
a reasonable prediction of the phase diagram of water), from SPC/E the use of 
Berendsen et al. correction to the vaporization enthalpy, and from TIP5P the use of 
the TMD as a target property. 
The resulting model, predicts quite nicely the orthobaric densities, 
critical temperature, surface tension, densities of the different solid phases 
of water, phase diagram, melting properties (with a reasonable prediction of 
the melting point about 20~K below the experimental value), TMD, isothermal 
compressibility, coefficient of thermal expansion, structure of water and ice. 
The properties analysed here cover a temperature range from 120 to 640 K and 
pressures up to 30 000 bar.

Obviously TIP4P/2005 is not the last word in water potentials and have some 
deficiencies.
It fails in the prediction of the vapour properties (vapour pressure, dew 
densities, critical pressure, second virial coefficient) and in the prediction
of the dielectric constant in the liquid phase. 
However, it really points out clearly the limits that can be achieved by rigid 
non-polarizable models. 
We believe that inclusion of polarizability within a TIP4P/2005-like model 
(with an adequate fitting of the parameters) would yield further improvement.
The extreme sensitivity of phase diagram to the water potential model may help 
in developing new potential models for water. 
Now that it is possible to evaluate phase diagrams (both vapour-liquid and 
fluid-solid) or the TMD on relatively routine basis, it sounds a good idea to 
extend the incorporation of the phase diagram and the room pressure isobar
as target properties in the fitting (and/or checking) of new water models. 
Although there may be a current of opinion suggesting that polarizable models
have not yet improved the performance of non-polarizable models, it is our 
belief that we have not worked hard enough to refine a reliable polarizable 
model for water. 
For the time being, TIP4P/2005 can be considered as a reliable and cheap model
of water offering good performance over a wide range of conditions. 
Since TIP4P/2005 is a minor modification of the model proposed by Bernal and
Fowler\cite{bernal33} in 1933, one may say that we have spend seventy five 
years in refining their parameters to finally obtain a reliable model of water
for the condensed phases. 
 
\acknowledgments
It is a pleasure to acknowledge to L. G. MacDowell, E.G.Noya, C. McBride, and R. G. 
Fernandez
for many helpful discussions and for their contribution to different parts of 
this work. 
Discussions with E.de Miguel (Huelva), A.Patrykiejew (Lublin), I.Nezbeda 
(Prague) are also gratefully acknowledged.
This work has been funded by grants FIS2007-66079-C02-01
from the DGI (Spain), S-0505/ESP/0229 from the CAM, MTKD-CT-2004-509249
from the European Union and 910570 from the UCM. 

\bibliographystyle{./apsrev}

\newpage 
\begin{table}
\begin{center}
\caption{Potential parameters of the water potential models.
The distance between the oxygen and hydrogen sites is d$_{OH}$.
The angle formed by hydrogen, oxygen, and the other hydrogen atom is denoted as 
H-O-H. 
The LJ site is located at the oxygen with parameters $\sigma$ and 
($\epsilon/k_{B}$). 
Proton charge q$_{H}$. 
All the models (except TIP5P) place the negative charge in a point M at a 
distance d$_{OM}$ from de oxygen along the H-O-H bisector. 
For TIP5P, d$_{OL}$ is the distance between the oxygen and the L sites placed 
at the lone electron pairs.}
\label{potentials}
\begin{tabular}{lccccccc}
\hline\hline
Model & d$_{OH}$ (\AA) & H-O-H & $\sigma$ (\AA) & ($\epsilon/k_{B}$) (K) & q$_H$ (e) & d$_{OM}$ (\AA) & d$_{OL}$ (\AA) \\
\hline
SPC & 1.0 & 109.47 & 3.1656 & 78.20 & 0.41 & 0 & - \\
SPC/E & 1.0 & 109.47 & 3.1656 & 78.20 & 0.423 & 0 & - \\
TIP3P & 0.9572 & 104.52 & 3.1506 & 76.54 & 0.417 & 0 & - \\
TIP4P & 0.9572 & 104.52 & 3.1540 & 78.02 & 0.52 & 0.15 & - \\
TIP4P/2005 & 0.9572 & 104.52 & 3.1589 & 93.2 & 0.5564 & 0.1546 & - \\
TIP5P & 0.9572 & 104.52 & 3.1200 & 80.51 & 0.241 & - & 0.70 \\
\hline
\end{tabular}

\end{center}
\end{table}

\newpage
\begin{table}
\caption{Dipole moment and components of the quadrupole moment.  
Debye units are used for the dipole moment $\mu$ while the components of the quadrupole moment are given in D$\cdot$\AA. $Q_T$ is defined as $Q_T=(Q_{xx}-Q_{yy})/2$.
The center of mass is used as origin, being the z axis that of the H-O-H 
bisector and the x axis located in the direction of the vector 
joining the two hydrogen atoms.}
\label{tabl_moments}
\begin{tabular}{lcccccc}
\hline\hline
 Model        &  $\mu$ & $Q_{xx}$ & $Q_{yy}$ & $Q_{zz}$ & Q$_T$ & $\mu/Q_T$\\
\hline
 SPC          &  2.274 \  &   2.12 \ &  -1.82 \ & -0.29 & 1.969 & 1.155 \\
 SPC/E        &  2.350 \  &   2.19 \ &  -1.88 \ & -0.30 & 2.035 & 1.155 \\
 TIP3P        &  2.350 \  &   1.76 \ &  -1.68 \ & -0.08 & 1.721 & 1.363 \\
 TIP4P        &  2.177 \  &   2.20 \ &  -2.09 \ & -0.11 & 2.147 & 1.014 \\
 TIP4P/2005   &  2.305 \  &   2.36 \ &  -2.23 \ & -0.13 & 2.297 & 1.004 \\
 TIP5P        &  2.290 \  &   1.65 \ &  -1.48 \ & -0.17 & 1.560 & 1.460 \\
 Gas(Expt.)   &  1.850 \  &   2.63 \ &  -2.50 \ & -0.13 & 2.565 & 0.721 \\
\hline
\end{tabular}
\end{table}

\newpage 

\begin{table}[!hbt]\centering
\caption{Vapor liquid equilibria for the TIP3P model as obtained from the 
computer simulations of this work.} 
\label{tip3p_VLE}
\begin{tabular}{cccc}
\hline\hline
T\,(K)& p\,(bar)& $\rho_{gas}$\,(g/cm$^3$)&$\rho_{liquid}$\,(g/cm$^3$)\\
\hline
550 & 81.32 & 0.074(3)  & 0.543(8)\\ 
545 &74.74  & 0.066(2) & 0.561(7)\\
540 &68.64  & 0.058(2)& 0.585(5)\\
530 &57.83  & 0.043(1) &0.613(5)\\
520 &48.44  & 0.0350(9) &0.643(7)\\
510 &40.48  & 0.0278(8) &0.668(4)\\
500 &33.58  & 0.0224(5) &0.697(4)\\
490 &27.62  & 0.0183(5) &0.720(4)\\
470 &18.30  & 0.0112(3) &0.760(5)\\
450 &11.72  & 0.0070(1) &0.799(3)\\
425 &6.313  & 0.00375(8) &0.836(3)\\
400 &3.142  & 0.00187(4) &0.873(2)\\
375 &1.412  & 0.00088(1)&0.904(2)\\
350 &0.5587 & 0.000361(6) &0.934(2)\\
325 &0.1890 & 0.000129(2) &0.959(2)\\
300 &0.05203& 0.0000380(4) &0.984(2) \\
275 &0.01101& 0.00000827(7) &1.004(2) \\
\hline
\end{tabular}
\end{table}
\newpage 
\vspace*{5cm}
\begin{table}
\caption{Critical properties of TIP3P, TIP4P, TIP4P/2005 and TIP5P.
         Values taken from Refs. [\protect\onlinecite{lisal01}] (TIP4P),
         [\protect\onlinecite{vega06}] (TIP4P/2005),
         [\protect\onlinecite{lisal02}] (TIP5P) and
         this work (TIP3P).}
\label{critical_properties}
\begin{tabular}{lccc}
\hline\hline
Model & $T_{c}$\,(K)& $p_{c}$\,(bar) & $\rho_c$\,(g/cm$^3$) \\
\hline
TIP3P         &  578      &  126   & 0.272 \\
TIP4P         &  588      &  149   & 0.315 \\
TIP4P/2005    &  640      &  146   & 0.31  \\
TIP5P         &  521      &   86   & 0.337 \\ 
Experiment    &  647.1    & 220.64 & 0.322         \\
\hline
\end{tabular}
\end{table}

\newpage 
\vspace*{4cm}
\begin{table}
\caption{Surface tension (in $mN/m$) for the TIP3P model of water at 
         different temperatures as obtained from computer simulation. 
         $\rho_l$ and $\rho_v$ are the densities (in g/cm$^3$) of the liquid 
         and vapour phases at coexistence. 
         $\bar{p}_{\rm N}$ and $\bar{p}_{\rm T}$ are the macroscopic values 
         of the normal and tangential components of the pressure tensor 
         (in bar units). 
         $t$ is the thickness of the vapour-liquid interface (in \AA). 
         $\gamma^*_{\rm v}$ and $\gamma^*_{\rm ta}$ are the values of the 
         surface tension obtained from the virial route and the test-area 
         method, respectively, without including long-range corrections.
         $\gamma_{\rm v}$ and $\gamma_{\rm ta}$ are the corresponding values 
         of the surface tension, but including long-range corrections which 
         correct for the truncation of the LJ part of the potential at 
         $r_{c}=13 $~{\AA} (see Ref.~[\onlinecite{vega07a}] for details). 
         The values of the surface tension as estimated from this work 
         ($\gamma_{\rm sim}$) correspond to the arithmetic average
         ($\gamma_{\rm v} + \gamma_{\rm ta})/2$. 
         $\gamma_{\rm exp}$ are the experimental values of the surface tension.}
\bigskip
\label{st_tip3p}
\begin{tabular}{cccccccccccc}
\hline\hline
$T$(K)& $\rho_l$ & $\rho_v$ & $\bar{p}_{\rm N}$ & $\bar{p}_{\rm T}$ & 
$\gamma^*_{\rm v}$ & $\gamma^*_{\rm ta} $ & $t$  & 
$\gamma_{\rm v}$   & $\gamma_{\rm ta}$ & $\gamma_{\rm sim}$ &  
$\gamma_{\rm exp}$ \\
\hline
 300.&  0.980&  0.000024&   -0.11&   -98.44&  49.2&  49.8&   3.87&  52.4& 52.2& 52.3(2.2)&  71.73  \\
 350.&  0.930&  0.00035&    0.22&   -80.63&  40.4& 40.7&  4.91&  43.3&  43.1&  43.2(2.0)&  63.22  \\
 400.&  0.867&  0.0018&    2.86&   -61.64&  32.3&  32.9&   6.12&  34.7&  34.5&  34.6(1.8)&  53.33  \\
 450.&  0.790&  0.0069&   11.95&   -33.81&  22.9&  23.3&   7.82&  24.8&  24.6&  24.7(1.5)&  42.88  \\
 500.&  0.689&  0.025&   33.01&     7.61&  12.7&  12.0&  11.53&  13.9&  13.8&  13.9(1.8)&  31.61  \\
\hline
\end{tabular}
\end{table}

\newpage 
\begin{center}
\begin{table}
\caption{Densities and residual internal energies of the different ices phases 
from NpT simulations for TIP3P and SPC.
The experimental data of ices are taken from Ref.~
[\onlinecite{bookPhysIce}], except those for ice VII (Ref.~
[\onlinecite{N_1987_330_00737_nolotengo}])
and ice II (Ref.~[\onlinecite{fortes_jac05}]).
The numbers in parenthesis correspond to the cases where the ices are not 
mechanically stable and melt into liquid water (the reported densities and
internal energies then correspond to those of the liquid).}
\label{density_ices}
\begin{tabular}{lccccccc}
\hline\hline
Phase &T\,(K)&p\,(bar)&
          \multicolumn{3}{c}{$\rho$ (g/cm$^3$)} &
                                     \multicolumn{2}{c}{U\,(kcal/mol)} \\
     &     &       & Exptl.&  SPC  & TIP3P &  SPC   & TIP3P \\
\hline
Liquid&300 &   1   & 0.996 & 0.975 & 0.982 &  -9.96 & -9.66 \\
Ih   & 250 &   0   & 0.920 & 0.923 &(1.021)& -11.72 &(-10.23)\\
Ic   &  78 &   0   & 0.931 & 0.951 & 0.959 & -13.04 & -12.49 \\
II   & 123 &   0   & 1.190 & 1.219 & 1.219 & -12.94 & -12.66 \\
III  & 250 &  2800 & 1.165 & 1.150 &(1.130)& -11.32 &(-10.34)\\
IV   & 110 &   0   & 1.272 & 1.298 & 1.286 & -12.23 & -11.75 \\
V    & 223 &  5300 & 1.283 & 1.270 &(1.226)& -11.60 &(-10.73)\\
VI   & 225 & 11000 & 1.373 & 1.379 & 1.366 & -11.47 & -10.91 \\
VII  & 300 &100000 & 1.880 & 1.809 & 1.826 &  -8.46 &  -8.16 \\
VIII &  10 & 24000 & 1.628 & 1.661 & 1.683 & -11.41 & -11.02 \\
IX   & 165 &  2800 & 1.194 & 1.198 & 1.194 & -12.55 & -12.16 \\
XII  & 260 &  5000 & 1.292 &(1.216)&(1.183)&(-10.78)&(-10.26)\\
XI   &  5  &   0   & 0.934 & 0.967 & 0.972 & -13.54 & -13.01 \\
XIII &  80 &   1   & 1.251 & 1.282 & 1.289 & -12.86 & -12.48 \\
XIV  &  80 &   1   & 1.294 & 1.33  & 1.338 & -12.65 & -12.22 \\
\hline
\end{tabular}
\end{table}
\end{center}

\begin{table*}[!]
\caption{
Melting properties of ice I$_h$ at $p=1$~bar for different models.
$T_{m}$, melting temperatures; 
$\rho_{l}$ and $\rho_{I_{h}}$, coexistence densities of liquid water and ice;
$\Delta$ $H_m$, melting enthalpy; 
$dp/dT$, slope of the coexistence curve (between ice Ih and water at the normal 
melting temperature of the model).
We also include for comparison the the ratio $T_{m}/T_{c}$.
Melting properties taken from Refs.~[\onlinecite{vegafilosofico}]
(TIP3P, SPC, SPC/E, TIP4P) and [\onlinecite{abascal05a}] (TIP4P/2005).
}
\label{propiedades_melting}
\begin{tabular}{lcccccccccc}
\hline \hline
Model
& TIP3P&\ SPC\ & SPC/E & TIP4P &TIP4P/2005& TIP5P& Exptl \\
\hline
$T_{m}$(K)
& 146& 190 &  215  & 232 &  252 &  274 &  273.15\\
$\rho_{l}$(g/cm$^3$)
& 1.017& 0.991 & 1.011 & 1.002 &  0.993 &  0.987 &  0.999\\
$\rho_{Ih}$(g/cm$^3$)
& 0.947& 0.934 & 0.950 & 0.940 &  0.921 &  0.967 &  0.917\\
$\Delta H_m$(Kcal/mol)
& 0.30 & 0.58  & 0.74  &  1.05 &  1.16  &  1.75  &  1.44 \\
$dp/dT$(bar/K)
&  -66 & -115  & -126  & -160  &  -135  &  -708  &  -137 \\
$T_{m}/T_{c}$
& 0.25 & 0.321 & 0.337 & 0.394 &  0.394 & 0.525  &  0.422\\
\hline \hline
\end{tabular}

\end{table*}

\begin{table}
\vspace*{5cm}
\caption{Temperature at which the maximum in density occurs $T_{TMD}$ (at room 
pressure) for different water models. 
The value of the coefficient of thermal expansion $\alpha$ and of the isothermal
compressibility $\kappa_T$ at room temperature and pressure are also given.  
The values of $\alpha$ and $\kappa_T$ for TIP3P, TIP4P and TIP5P as reported in 
Ref.\onlinecite{jorgensenreview}. 
The values for TIP4P/2005 were taken from Ref.\cite{abascal05b}.
The value of the maximum in density as given in Ref.\cite{vega05b}, except
those of the TIP5P that were taken from the original reference.\cite{mahoney00}.}
\label{tmd_alpha_kappa}
\begin{tabular}{cccc}
\hline\hline
  Model & TMD (K)&$10^{5}\kappa_{T}$ (MPa$^{-1}$)
                  &$10^{5}\alpha$ (K$^{-1}$)\\
\hline
  TIP3P   & 182  &   64    &  92 \\
  TIP4P   & 253  &   59    &  44 \\
  TIP5P   & 277  &   41    &  63 \\
TIP4P/2005& 278  &   46    &  28 \\
  Expt    & 277  &   45.3  & 25.6 \\
\hline
\end{tabular}
\end{table}

\begin{table}
\begin{center}
\caption{Self-diffusion coefficient 10$^9D$ (m$^2$/s) for liquid water at 1 bar
as a function of temperature.}
\label{tablaD}
\begin{tabular}{lccccc}
\hline \hline
T (K)& TIP3P & TIP4P & TIP5P & TIP4P/2005 & Exp.\\
\hline
278 & 3.71 & 2.08 & 1.11 & 1.27 & 1.313\\
288 & 4.34 & 2.71 & 1.74 & 1.57 & 1.777\\
298 & 5.51 & 3.22 & 2.77 & 2.07 & 2.299\\
308 & 6.21 & 4.12 & 3.68 & 2.60 & 2.919\\
318 & 6.32 & 4.90 & 4.81 & 3.07 & 3.575\\
\hline
\end{tabular}
\end{center}
\end{table}

\begin{table}
\caption{Computed static dielectric constant $\varepsilon$ at 298 K and 1 bar.}
\begin{tabular}{lc}
\hline\hline
Model & $\varepsilon$ \\
\hline
TIP3P & 94 \\
TIP4P & 50 \\
TIP4P/2005 & 58 \\
TIP5P & 91 \\
Exp. & 78.4 \\
\hline
\end{tabular}
\label{que_pasa}
\end{table}

\newpage 
\vspace*{5cm}
\begin{table}
\begin{center}
\caption{Scores obtained by each model for the ten properties considered in 
this work.}
\label{final_exam}
\begin{tabular}{lcccc}
\hline\hline
Property                &TIP3P&TIP4P&TIP4P/2005&TIP5P \\
\hline
1. VLE, $T_{c}$          &  1  &  2  &   3   &  0 \\
2. Surface tension       &  1  &  2  &   3   &  0 \\ 
3. $\rho$ ices           &  0  &  2  &   3   &  1 \\ 
4. Phase diagram         &  0  &  2  &   3   &  1 \\
5. $T_{m}$ melting prop. &  0  &  1  &  2.5  & 2.5 \\
6. $T_{TMD}$, $\alpha$, $\kappa_{T}$
                         &  0  &  1  &   3   &  2 \\
7. Structure             &  0  &  1  &  2.5  & 2.5 \\
8. EOS (high p)          &  2  &  1  &   3   &  0 \\
9. D                     &  0  &  1  &   3   &  2 \\
10. $\epsilon$           &  2  &  0  &   1   &  3 \\
Total                    &  6  & 13  &  27   & 14 \\ 
\hline
\end{tabular}
\end{center}
\end{table}

\newpage 
\vspace*{5cm}

\newpage 
\vspace*{5cm}

\newpage 
\vspace*{5cm}

\newpage 
\vspace*{5cm}

\vspace*{2cm}


\vspace*{2cm}

\newpage 
\vspace*{5cm}

\vspace*{1cm}

\newpage

\vspace*{2cm}

\newpage
\end{document}